\documentclass{ws-ijmpa}

\begin{document}

\markboth{Julio Oliva, David Tempo and Ricardo Troncoso}
{Static spherically symmetric solutions of three-dimensional conformal gravity}

%
\catchline{}{}{}{}{}
%

\title{Static spherically symmetric solutions
\\ for conformal gravity in three dimensions}

\author{Julio Oliva}

\address{Centro de Estudios Cient\'{\i}ficos (CECS), Casilla 1469, Valdivia, Chile.}

\author{David Tempo}

\address{Centro de Estudios Cient\'{\i}ficos (CECS), Casilla 1469, Valdivia, Chile. \\
Departamento de F\'{\i}sica, Universidad de Concepci\'{o}n, Casilla,
160-C, Concepci\'{o}n, Chile.}

\author{Ricardo Troncoso}

\address{Centro de Estudios Cient\'{\i}ficos (CECS), Casilla 1469, Valdivia, Chile.\\
Centro de Ingenier\'{\i}a de la Innovaci\'{o}n del CECS (CIN), Valdivia, Chile.}

\maketitle

\begin{history}
\received{30 September 2008}
\end{history}

\begin{abstract}
Static spherically symmetric solutions for conformal gravity in three dimensions are found. Black holes and wormholes are included within this class. Asymptotically the black holes are spacetimes of arbitrary constant curvature, and they are conformally related to the matching of different solutions of constant curvature by means of an improper conformal transformation. The wormholes can be constructed from suitable identifications of a static universe of negative spatial curvature, and it is shown that they correspond to the conformal matching of two black hole solutions with the same mass.

\end{abstract}

\ccode{PACS numbers: 04.20.Jb, 04.50.Kd}

\section{Introduction}	
The lack of propagating degrees of freedom for General Relativity (GR) in three dimensions stems from the fact that the solutions must be spacetimes of constant curvature (see e.g. Ref. \refcite{C}). Nonetheless, for negative cosmological constant, nontrivial solutions including black holes can be found from suitable identifications of anti-de Sitter (AdS) spacetime\cite{BHTZ}. Besides, few is known about conformal gravity in three dimensions, whose field equations correspond to the vanishing of the Cotton tensor,
\begin{equation}
C_{\hspace{0.05in}\nu}^{\mu}=\epsilon^{\kappa\sigma\mu}\nabla_{\kappa}\left(
R_{\sigma\nu}-\frac{1}{4}g_{\sigma\nu}R\right)  =0,\label{fieldequations}%
\end{equation}
which are fulfilled if and only if the spacetime metric is locally conformally flat. Exact solutions for this theory have been recently explored in Refs. \refcite{Z1,Z2}. The purpose of this paper is showing that, since the field equations (\ref{fieldequations}) are conformal invariant, interesting nontrivial solutions can be found not only from suitable identifications, but also from improper conformal transformations of maximally symmetric spacetimes. It is worth pointing out that wormholes as well as asymptotically locally flat or de Sitter (dS) black hole solutions arise within this new set. This can be seen as follows: It is possible to choose the gauge such that the static spherically symmetric solution of (\ref{fieldequations}) reads%
\begin{equation}
ds^{2}=-\left(  ar^{2}+br+c\right)
dt^{2}+\frac{dr^{2}}{ ar^2+br+c
}+r^{2}d\phi^{2},\label{final}%
\end{equation}
where $a,b$ and $c$ are integration constants. These solutions are asymptotically of constant curvature $-a$, which by means of a trivial (proper) global conformal transformation can be rescaled to $\pm1$ or zero. For vanishing $b$ the metric (\ref{final}) has constant curvature, and it reduces to the usual solution of standard GR. Thus, switching on the constant $b$ relaxes the asymptotic behavior of the metric as compared with the one of GR, enlarging the space of allowed solutions. Indeed, for $b\neq0$ the Ricci scalar is given by $R=-6a-2b r^{-1}$, which is singular at the origin. Depending on the value of the integration constants, this singularity could be surrounded by one or two horizons.

In the case of vanishing $a$, for $b>0$ and $c<0$, the metric (\ref{final}) describes an asymptotically locally flat black hole with a spacelike singularity at the origin surrounded by an event horizon located at $r=r_{+}:=-cb^{-1}$. Its causal structure coincides with the one of the Schwarzschild black hole (see Fig.1 C.1).

The case $a=-1$, corresponds to an asymptotically dS black hole with a spacelike singularity at the origin enclosed by event and cosmological horizons located at $r_{+}$ and $r_{++}$, respectively, provided $b=r_{+}+r_{++}$ and $c=-r_{+}r_{++}$. As shown in Fig.1 B.1, this black hole shares the same causal structure with the Schwarzschild-dS metric.

It is worth to remark that static black holes cannot be obtained from three-dimensional GR with non negative cosmological constant in vacuum.

Asymptotically AdS black holes are obtained for the case $a=1$. For $c>0$ and $b<0$ the curvature singularity is timelike and it is surrounded by a Cauchy and an event horizon located at $r_{-}$ and $r_{+}$, respectively, provided $b=-(r_{-}+r_{+})$ and $c=r_{-}r_{+}$. In this case the causal structure corresponds to the one of the Reissner-Nordstrom-AdS black hole (see Fig.1 A.1). The extremal case is obtained for $r_{+}=r_{-}$. For negative $c$ ($r_{-}<0$) the black hole possesses a spacelike curvature singularity at the origin, surrounded by a single event horizon located at $r_{+}$, and its causal structure reduces to the one of the Schwarzschild-AdS black hole, as it is depicted Fig.1 D.1.

The ``Schwarzschild gauge" is not the only option. Actually, a different gauge fixing leads to the following static spherically symmetric solution:
\begin{equation}
ds_{w}^{2}=-dt^{2}+dz^{2}+l_{0}^{2}\cosh^{2}\left(  z\right)  d\phi^{2},%
\label{wormholeansatz}%
\end{equation}
describing a wormhole with a neck of radius $l_{0}$ located at $z=0$. This spacetime is the product of the real line with a hyperbolic space identified along a boost ($R\times H_{2}/ \Gamma$), so that it connects two static universes of negative spatial curvature with unit radii, located at $z\rightarrow\pm \infty$. The causal structure of the wormhole (\ref{wormholeansatz}) coincides with the one of Minkowski spacetime in two dimensions, as it is depicted in Fig.1 E.1.

It can be shown that the black holes (\ref{final}) and the wormhole (\ref{wormholeansatz}) correspond to the
matching of different patches of constant curvature spacetimes \emph{at spatial infinities}, by means of improper conformal transformations. It is worth pointing out that the matching of different spaces through the boundary of their corresponding conformal compactifications cannot be performed in GR, since the proper distance, $ds_{E}^{2}$, to pass from one patch to the other diverges. However, in conformal gravity this kind of matching can be carried out by means of an improper conformal transformation, i.e., a local rescaling $\Omega^{2}$ that vanishes at the matching surfaces. In this way, the proper distance required to pass through a pair of points located at each side of the matching surface, given by $ds^{2}=\Omega^{2}ds_{E}^{2}$, becomes finite. Note that this procedure works in a way that is analogous to the one required to obtain black holes in $1+1$ dimensions (see e.g. \refcite{Witten,Katanaev}).

In the case of the wormhole (\ref{wormholeansatz}), the metric is conformally related to the matching of two independent patches covering the exterior region of BTZ black holes, where the horizon radius is related to the radius of the neck according to $\rho_{+}=l_{0}$. This is shown in Fig.1 E.2.

The class of black holes described by (\ref{final}) is conformally related to the matching of different independent patches of static spherically symmetric Einstein spaces of mass $M=-c$ with a cosmological constant of sign $sgn(c)$ (See Fig.1 A-D).

One may wonder about the possibility of generating new static spherically symmetric solutions of conformal gravity with an arbitrary number of horizons through the ``conformal matching" procedure described above. It is possible to show that the most general solution admits at most two horizons, since the attempt of including a third one necessarily introduces an additional curvature singularity at a finite radius between the second and the third horizon. Hence, once the region inside this singular shell is excised, one necessarily recovers a static spherically symmetric black hole possessing one or two horizons only, described by the metric (\ref{final}).
The solutions presented here can be extended to the rotating case\cite{OTT2}. The definition of suitable conserved charges and the black hole thermodynamics within this theory is an open problem.
\begin{figure}[pb]
\centerline{\psfig{file=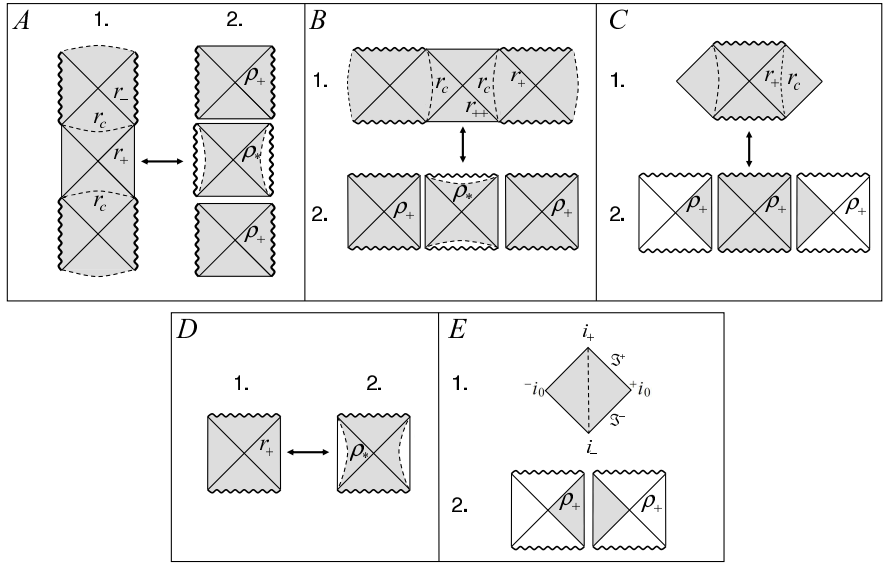,width=8.1cm,angle=-90}}
\caption{ The causal structure of the conformal gravity black holes (\ref{final}) and the wormhole (\ref{wormholeansatz}) is shown in (A-D).1 and E.1, respectively. These spacetimes can be constructed from the conformal matching of different patches of constant curvature spacetimes at the corresponding
spatial infinities, as depicted in (A-E).2. Each square diagram  appearing in Figs.(B-E).2 describes the causal structure of static BTZ black holes. In Fig. A.2 the square diagrams correspond to the causal structure of static solutions with negative mass and positive cosmological constant. In the figures A.2 and B.2, the region $\rho<\rho_{*}$ must be excised and for D.2 the excised region correspond to $\rho>\rho_{*}$. Here $r_{c}$ corresponds to the matching surface where the constant curvature spacetimes are smoothly glued; for the wormhole the matching surface is located at the neck ($z=0$). }
\end{figure}
\section*{Acknowledgments}
We thank the organizers of the meeting for their kind hospitality. Special thanks to M. Katanaev and R.E. Troncoso for very useful comments. This work was
partially funded by FONDECYT grants 1095098, 1061291, 1071125, 1085322, 3085043; D. Tempo thanks CONICYT and Escuela de Graduados (UdeC), for financial support. The Centro
de Estudios Cient\'{\i}ficos (CECS) is funded by the Chilean Government
through the Millennium Science Initiative and the Centers of Excellence Base
Financing Program of CONICYT. CECS is also supported by a group of private
companies which at present includes Antofagasta Minerals, Arauco, Empresas
CMPC, Indura, Naviera Ultragas and Telef\'{o}nica del Sur. CIN is funded by
CONICYT and the Gobierno Regional de Los R\'{\i}os.

\end{document}